\documentclass[10pt,conference]{IEEEtran}
\usepackage{times}

\usepackage{graphicx,subfigure}
\usepackage{amsmath}
\usepackage{dcolumn}
\usepackage[english]{babel}
\usepackage{amssymb}
\usepackage{url}                        
\usepackage[super,negative]{nth}        
\usepackage{icomma}                     
\renewcommand{\subsubsection}[1]{\vspace*{-1ex}\subsection*{\center{\textbf{#1}\vspace*{-2ex}}}}
\usepackage[ruled]{algorithm2e}
\usepackage{dcolumn}

\newcommand{\co}[2]{#1 \tiny{$\pm#2$}} 
 
\begin{document}

\title{Evaluating Mobility Pattern Space Routing\\for DTNs}

\author{
\authorblockN{J\'er\'emie Leguay,\authorrefmark{1}\authorrefmark{2} Timur Friedman,\authorrefmark{1} Vania Conan\authorrefmark{2}\\}
\authorblockA{\authorrefmark{1} Universit\'e Pierre et Marie Curie, Laboratoire LiP6--CNRS\\
\authorrefmark{2} Thales Communications
}
}

\maketitle

\begin{abstract}
Because a delay tolerant network (DTN) can often be partitioned,
the problem of routing is very challenging.
However, routing benefits considerably if one can take advantage 
of knowledge concerning node mobility.
This paper addresses this problem with a generic algorithm based on the use
of a high-dimensional Euclidean space, that we call MobySpace, constructed upon
nodes' mobility patterns. We provide here an analysis and the 
large scale evaluation of this routing scheme in the context of ambient networking by replaying
real mobility traces. The specific MobySpace evaluated is based on the frequency of visit
of nodes for each possible location.
We show that the MobySpace can achieve good performance 
compared to that of the other algorithms we implemented, 
especially when we perform routing on the nodes that have a high
connection time. We determine that the degree of homogeneity of mobility patterns of nodes 
has a high impact on routing. And finally, we study 
the ability of nodes to learn their own mobility patterns. 
\end{abstract}

\section{Introduction}\label{sec_introduction}

This paper addresses the problem of routing in delay tolerant networks 
(DTNs)~\cite{dtn_fall_sigcomm}. It evaluates a scheme, proposed in~\cite{wdtn}, 
that turns the problem of DTN routing into a problem of routing in a virtual space defined by
the mobility patterns of nodes. The earlier work tested the scheme with an entirely artificial scenario. 
By driving simulations with real mobility traces, in this paper we validate 
this routing scheme in the context of ambient networks. This paper also studies 
a number of important factors, such as the degree of homogeneity in the mobility of nodes, 
that impact routing performance.  Finally, the paper examines the ability of nodes to learn their own mobility,
which is important for the feasibility of such a scheme.

In one common DTN scenario, like the one we consider in this 
paper, nodes are mobile and have wireless
networking capabilities. They are able to communicate together only
when they are within transmission range. The network suffers from
frequent connectivity disruptions, making the topology intermittently
and partially connected. This means that there is a very low
probability that an end-to-end path exists between a given pair of
nodes at a given time.  End-to-end paths can exist temporarily, or may
sometimes never exist, with only partial paths emerging.  Due to these
disruptions, regular ad-hoc networking approaches to routing and
transport do not hold, and new solutions must be proposed.

The Delay Tolerant Network Research Group (DTNRG)~\cite{dtnrg} has
proposed an architecture~\cite{draftirtf} to support messaging that
may be used by delay tolerant applications in such a context. The
architecture consists mainly of the addition of an overlay, called the
bundle layer, above a network's transport layer. Messages transferred
in DTNs are called bundles. They are transferred in an atomic fashion
between nodes using a transport protocol that ensures node-to-node
reliability. These messages can be of any size. Nodes are assumed to
have buffers in which they can store the bundles.

Routing is one of the very challenging open issues in DTNs, as
mentioned by Jain et al.~\cite{routingdtn}. Indeed, since the network
suffers from connectivity problems, MANET~\cite{rfc2501} routing
algorithms such as OLSR, based on the spreading of control
information, or AODV, which is on-demand, fail to achieve routing.
Different approaches have to be found.

The problem of routing in DTNs is not trivial. Epidemic routing~\cite{vahdat00epidemic}
studied by Vahdat et al.\ is a possible solution when nothing is known about
the behavior of nodes. Since it leads to buffer overloads and
inefficient use of transmission media, one would prefer to limit
bundle duplication and instead use routing heuristics that can take
advantage of the context. To move in such a direction, the DTN
architecture defines several types of contacts: \emph{scheduled},
\emph{opportunistic}, and \emph{predicted}. \emph{Scheduled} contacts can exist,
for instance, between a base station somewhere on earth and a low
earth orbiting relay satellite.  \emph{Opportunistic} contacts are
created simply by the presence of two entities at the same place, in a
meeting that was neither scheduled nor predicted. Finally,
\emph{predicted} contacts are also not scheduled, but predictions of
their existence can be made by analyzing previous observations. 

The study presented in this paper relies also on contacts that
can be characterized as predicted, but the underlying concept is a more
generic abstraction compared to previous work, being able to capture
the interesting properties of major mobility patterns for routing.

The main contribution of this paper is the validation of a routing
scheme for DTNs that uses the formalism of a high-dimensional Euclidean
space based on nodes' mobility patterns.  We show the feasibility of
this concept through an example in which each dimension represents
the probability for a node to be found in a particular location.
We conduct simulations by replaying mobility traces to analyse the
feasibility and comparative performance of such a scheme.

The rest of this paper is structured as follows.  
Sec.~\ref{sec_mobyspace} describes the general concept of the mobility 
pattern based routing scheme, called the MobySpace. Sec.~\ref{sec_freq_of_visit_MobySpace} 
presents the specific MobySpace we have considered for the evaluation.
Sec.~\ref{sec_simus} presents the simulation results.
Sec.~\ref{related} provides an overview of related work concerning routing
in DTNs.
Sec.~\ref{sec_conclu} concludes the paper, discussing directions for
future work.

\section{MobySpace: a Mobility Pattern Space}\label{sec_mobyspace}

Two people having similar mobility patterns are more likely to meet 
each other, thus to be able to communicate. Based on this basic principle, we 
propose in~\cite{wdtn} to use the formalism of a Euclidean virtual space, that we call \emph{MobySpace},
as a tool to help nodes make routing decisions.  
These decisions rely on the notion that a node is a good
candidate for taking custody of a bundle if it has a mobility pattern
similar to that of the bundle's destination.
Routing is done by forwarding bundles toward nodes that have mobility
patterns that are more and more similar to the mobility pattern of the
destination. Since in the MobySpace, the mobility pattern of a node
provides its coordinates, called its \emph{MobyPoint}, routing is done by
forwarding bundles toward nodes that have their MobyPoint closer and
closer to the MobyPoint of the destination.  

In this section, we describe manners in which mobility patterns can be
characterized and the ways these patterns can be managed by the nodes, and we 
discuss possible limits and issues surrounding the overall concept.  

\subsection{Mobility pattern characterization}

Since the mobility pattern of a node provides its coordinates in the MobySpace,
the way in which these patterns are characterized determines the way the virtual
space is constructed. 

The definition of a mobility pattern
sets the number and the type of the dimensions of the specific MobySpace.
Many methods could be employed to describe a mobility pattern, but 
some requirements must be satisfied. We want mobility patterns to be 
simple to measure in order to keep them 
computationally inexpensive and to reduce the overhead associated with exchanging them between nodes. 
Furthermore, they must be relevant to routing, by helping nodes 
to take efficient routing decisions. 

A mobility pattern could be based, for instance, upon historic information regarding
contacts that the node has already had. A recent study~\cite{pocket_wdtn} by 
Hui et al. has shown the interest of such mobility patterns. It
highlighted that contacts between people at the Infocom 2005 conference
follow power-laws in term of their duration.
If we want to route a bundle from one node to another, we have an
interest in considering information on these contacts. Intuitively, it could be very 
efficient to transmit a bundle to a relay that frequently encounters the destination. 
A MobySpace based on this kind of pattern would be as follows.
Each possible contact is an axis, and the distance
along that axis indicates a measure of the probability of contact.
Two nodes that have a similar set of contacts that they see with
similar frequencies are close in this space, whereas nodes that have
very different sets of contacts, or that see the same contacts but
with very different frequencies, are far from each other. It seems
reasonable that one would wish to pass a bundle to a node that is as
close as possible to the destination in this space, because this
should improve the probability that it will eventually reach the
destination.

We might wish to consider an alternative space in which there is a more 
limited number of axes.  If nodes' visits to particular
locations can be tracked, then the mobility pattern of a node can be
described by its visits to these locations.  In this scenario, each
axis represents a location, and the distance along the axis represents
the probability of finding a node at that location.  We can imagine
that nodes that have similar probabilities of visiting a similar set of
locations are more likely to encounter each other than nodes that are
very different in these respects.  

Prior work~\cite{routingdtn}, has demonstrated the interest of capturing 
temporal information as well.  It is well known that network usage patterns
follow diurnal and weekly cycles.  We could easily imagine two nodes
that visit the same locations with the same frequencies, but on
different days of the week. This kind of desynchronisation could arise for instance
in a campus at the scale of the hour if we consider two users each having a course 
in the same lecture hall the same day but not at the same time.
Even so, it still might make sense to
route to one node in order to reach the other, especially if there
is a relay node at the commonly visited location. We can imagine ways in which the
dimensional representation could capture temporal information as
well. For instance, visit patterns could be translated into the
frequential domain.  A node's visits to a location could be
represented by a point on a frequency axis, capturing the dominant
frequency of visits, and a point on a phase axis, as well as a point
on the axis already described, that represents the overall
probability of visiting the location.

The evaluation and the comparison of the different kind of mobility 
patterns are kept for further studies. 
In Secs.~\ref{sec_freq_of_visit_MobySpace} and \ref{sec_simus}, we test a MobySpace 
based on the frequency with which
nodes find themselves in certain locations

\subsection{Mobility pattern acquisition}

A node in the network has to determine its coordinates in the MobySpace,  
the ones of the nodes it meets, and the ones of the destinations of the bundles 
it carries, in order to take appropriate routing decisions. Two problems arise:
how does a node learn its own mobility pattern, and how does a node 
learn those of the others? 

There are several ways a node can learn its own mobility pattern.
First, a node can learn its mobility pattern by observing 
its environment, e.g., by studying its contacts or its frequency of visits
to different locations. If the node requires information about its current position, we 
can assume that particular tags are attached to each location. 
Alternatively, we can imagine that nodes are able to interrogate an exiting 
infrastructure to obtain these patterns. This infrastructure would act as a passive 
monitoring tool for pattern calculation. The system can be accessible anywhere 
in a wireless or in a wired fashion or it can 
be located at certain places.

Similarly, there are several ways that a node can learn the mobility patterns of others.
First, these mobility
patterns can be spread in an epidemic fashion. Meaning that they do not need to 
be retransmitted by nodes in the network if we assume that there are always nodes
present in the network to make the information remain available to new
nodes, and if we assume that the mobility patterns of nodes are stable. 
Nodes could also spread just the most significant coordinates
of their mobility patterns to reduce buffer occupancy and network
resource consumption.
Finally, we can also imagine that nodes drop off their mobility patterns in repositories
placed at strategic locations, and at the same time they update 
their knowledge with the content available at the repositories.

\subsection{Mobility pattern usage}

As mentioned in the introduction, the mobility pattern of a node determines its coordinates 
in the MobySpace, i.e., the position of its associated MobyPoint. The basic idea
is that bundles are forwarded to nodes having mobility patterns more and more 
similar to that of the destination. This means that in the MobySpace, we route
bundles to nodes that have MobyPoints closer and closer to that of the destination.
Let's consider that the node $A$ wants to route a bundle to $B$, while only the node 
$C$ is in its neighborhood. It has two possibilities: keep the bundle or transmit the 
bundle to $C$. The MobyPoint of $A$ is at a certain distance from the MobyPoint of $B$. 
If the MobyPoint of $C$ is closer to the MobyPoint of $B$, then node $A$ will decide to transmit to $C$.
Otherwise, it will keep the bundle.

Formally, let $U$ be the set of all nodes and $L$ be the set of all
locations.  The MobyPoint for a node $k \in U$ is a point in an
$n$-dimensional space, where $n = |L|$.  We write $m_k=(c_{1_k}, ...,
c_{n_k})$ for the MobyPoint of node $k$.  The distance between two
MobyPoints is written $d(m_i,m_j)$.

At a point in time, $t$, the node $k$ will have a set of directly
connected neighbors, which we write as $W_{k}(t) \subseteq U$.
$W_{k}^{+}(t)=W_{k}(t) \cup \{k\}$ is the augmented neighborhood that contains $k$.
MobySpace routing consists of either choosing one of these neighbors to
receive the bundle or deciding to keep the bundle. The routing function, which we call $f$,
chooses the neighbor that is closest to the destination. The decision for node $k$ 
when sending a bundle to $b$ is taken by applying the function $f$: 

\begin{multline} 
f(W_{k}^{+}(t),b) =\\
\begin{cases}
\text{$b$ if $b \subset W_{k}(t)$, else}  \\
i \in W_{k}^{+}(t) : d(m_i,m_b) = \min_ {j \in W_{k}^{+}(t)} d(m_j,m_b)
\end{cases}
\end{multline}

The choice of the distance function $d$ used in the routing decision process is important. 
One straightforward choice is Euclidean distance. Examples of other distance functions 
can be found in~\cite{wdtn}. We leave their comparison to future work.

\subsection{Possible limits and issues}

DTN routing in a contact space or a mobility space is based on the
assumption that there will be regularities in the contacts that nodes
have, or in their choices of locations to visit.  There is always the
possibility that we may encounter mobility patterns similar to the
ones observed with random mobility models.  The efficiency of the
virtual space as a tool may be limited if nodes change
their habits too rapidly.

Some problems could occur even if nodes have well defined mobility
patterns, but their existence and nature may depend on the
particularities of the space.
For instance, in the Euclidean space, a bundle may reach a local
maximum if a node has a mobility pattern that is the most similar in
the local neighborhood to the destination node's mobility pattern, but
is not sufficient for one reason or another to achieve the delivery.
In the second type of space, where each dimension represents a
location, it can happen if nodes visit similar places, but for timing
reasons, such as being on opposite diurnal cycles, they never meet.
This kind of user behavior has been observed by Henderson et al.~\cite{dartmount} and 
Hui et al.~\cite{pocket_wdtn}. 

The Euclidean spaces that we have discussed here are finite in 
terms of number of dimensions, but in practice the number of dimensions might be unbounded.
This is the case, for instance, 
in the space we use as a case study in Sec.~\ref{sec_freq_of_visit_MobySpace}.
Additional mechanisms must be found to allow this.

Finally, the routing scheme presented here is based on each node forwarding just a single copy of a bundle, which may be 
a problem in case of node failure or nodes leaving the system for extended
periods of time. One may wish to introduce some redundancy into MobySpace routing.
For instance, a node can be allowed to transmit a bundle up to $T$ times if,
after the first transmission, it meets other nodes having mobility patterns 
even more similar to that of the destination within a period $P$.

\section{Frequency of visit based MobySpace}\label{sec_freq_of_visit_MobySpace}

To evaluate the routing scheme based on MobySpace,
we used a simple kind of space that we describe in the first part of this
section. The second part introduces the mobility data we replayed for 
the evaluation.

\subsection{Description}

The frequency of visit based MobySpace we evaluate works as follow. 
For each of the nodes, there is a well defined
probability of finding that node at each of the $n$ locations.  This
set of probabilities is a node's mobility pattern, and is described by
a MobyPoint in an $n$ dimensional MobySpace. The coordinate of a node 
for the axis $i$ is its probability of visit for the location $i$. 
All the MobyPoints are placed on a hyperplane since we have:

\begin{equation}
\textrm{for any point $m_{i} = (c_{1_i}, ..., c_{n_i})$, $\sum_ {k=1}^{n} c_{k_i} = 1$}
\end{equation}

One major motivation for this kind of space is the fact
that the distribution of the probabilities of visit to locations
follows generally a power law which can be used efficiently 
for routing. 

Indeed, recent studies of the mobility of students in a campus
\cite{ucsd,dartmount} or corporate users \cite{balazinska2003characterizing}, 
equipped with PDAs or laptops able
to be connected to wireless access networks, show that they follow
common mobility patterns.  They show that significant aspects of the
behavior can be characterized by power law distributions.
Specifically, the session durations and the frequencies of the places
visited by users follow power laws.  This means that users typically
visit a few access points frequently while visiting the others rarely,
and that users may stay at few locations for long periods while
visiting the others for very short periods. Henderson et al.
observed~\cite{dartmount} that $50$\% of users studied spent $62$\% of
their time attached to a single access point, and this proportion
decreased exponentially.  

Regarding the distance function, we choose a straightforward one, 
the Euclidean distance:

\begin{equation}
d(m_i,m_j)=\sqrt{\sum_ {k=1}^{n} \left(c_{k_i}-c_{k_j} \right)^2}
\end{equation}

\subsection{Real mobility data used}\label{sec_data}

There has been considerable growth in the number of small devices people carry every day,
such as cell phones, PDAs, music players, and game consoles. The variety of their
different networking capabilities allows us to envisage new applications,
such as distributed databases, content delivery systems, or
self organizing peer to peer networks. We can imagine that such spontaneous and autonomous networks spring up
around the movement of people in campus or corporate environments.
Contextual applications, services, or basic applications like text messaging could
take advantage of such an infrastructure. These scenarios are studied within the 
framework of delay tolerant networks.

\begin{figure}[!h]
\begin{center}
\includegraphics[width=6cm]{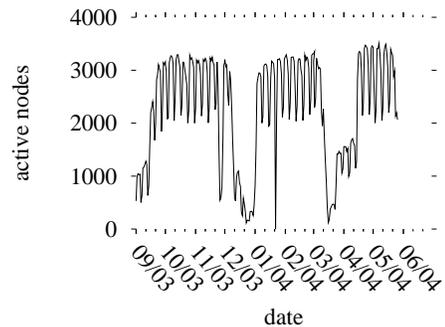}
\caption{\label{fig_activenodes} Number of actives users per day (from 1 September 2003 to 1 June 2004).}
\end{center}
\end{figure}

For the purpose of this study, we sought real mobility traces that ressembled what one might find 
in an ambient network environment. Since there are very few traces of this kind, we  
chose data that tracks mobile users in a campus setting.  
We used the mobility data collected on the {W}i-{F}i 
campus network of Dartmouth College~\cite{dartmount}.
Jones et al.~\cite{jones_wdtn} have recently used the traces in a similar way. 
The Dartmouth data is the most extensive data collection available that covers a 
large wireless access network. The network 
is composed of about $550$ access points (APs), the number of different wireless cards (MAC 
addresses) seen by the network is about $13,000$ and the data have been collected between the years 
$2001$ and $2004$. The network covers the college's academic 
buildings, the library, the sport infrastructures, the administrative buildings 
and the student residences. Users are equipped with devices such as PDAs, 
laptops, and phones that support voice over IP (VoIP). The majority of the end 
users are students, who make intensive use of the network, especially since many of them 
are required to own a laptop. Fig.~\ref{fig_activenodes} illustrates the usage levels by showing the 
evolution of the number of active nodes in the network per day.

The data we analysed track users' sessions in the wireless
network. 
These data have been pre-processed by Song et al.\ in their prior work~\cite{song:predict}
on mobility prediction.
The traces show the time at which a node associates or dissociates from an
access point. Data were collected by a central server with the Syslog~\cite{rfc3164} protocol. 
It could happen that a node does not send  
a dissociation message, or that a Syslog UDP message is lost, in which case a 
session is considered finished after 30 minutes of inactivity.

For our study, each access point represents 
a location. We assume that two nodes (represented as networking cards in 
the data) are assumed to be able to communicate with a low range device (using Bluetooth for instance), 
if they are attached at the same time to the same AP. This assumption
is somewhat artificial as nodes that are attached to 
two different APs that close to each other might be able to communicate directly.
Similarly, two nodes connected at the same AP might be out of range of each other. 
Nonetheless, this is the best approximation we can make with the data at hand.

\section{Simulation results}\label{sec_simus}

This section presents the manner in which we evaluated the routing scheme that uses 
a frequency of visit based MobySpace, and the results we obtained.
Since we performed the simulations using a subset of $45$ days of mobility data, 
we first describe the properties of the traces collected during that period.

\subsection{Mobility traces}\label{sec_datas_for_simulations}

\begin{figure}[!htb]
\centering
\subfigure[Active days]{\label{activedays}\includegraphics[width=4cm]{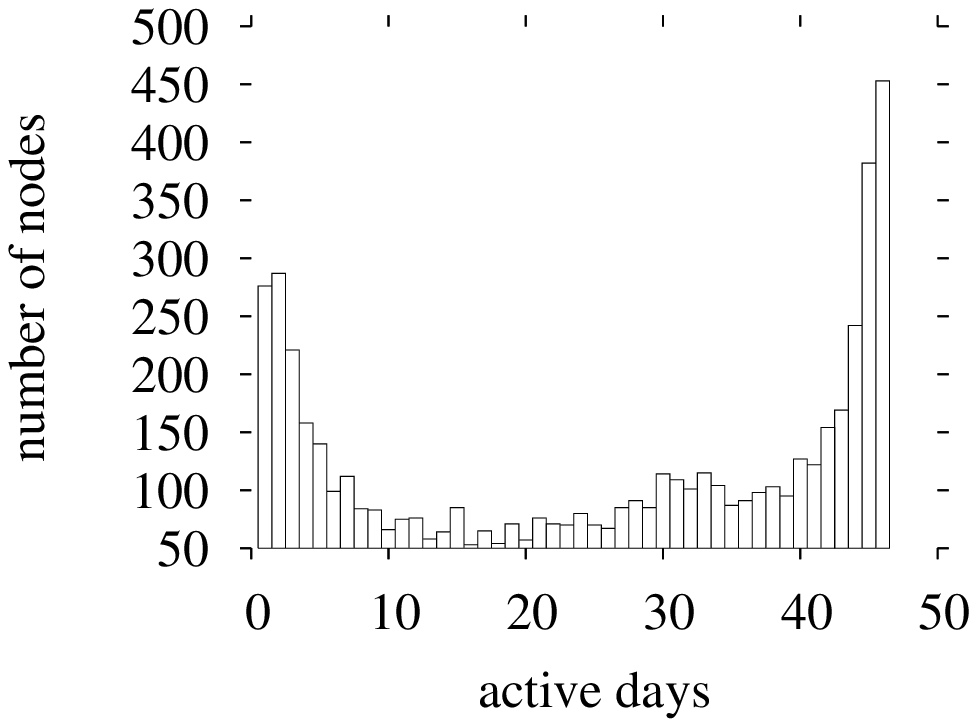}}
\subfigure[Active nodes per day]{\label{activenodes_perday}\includegraphics[width=4cm]{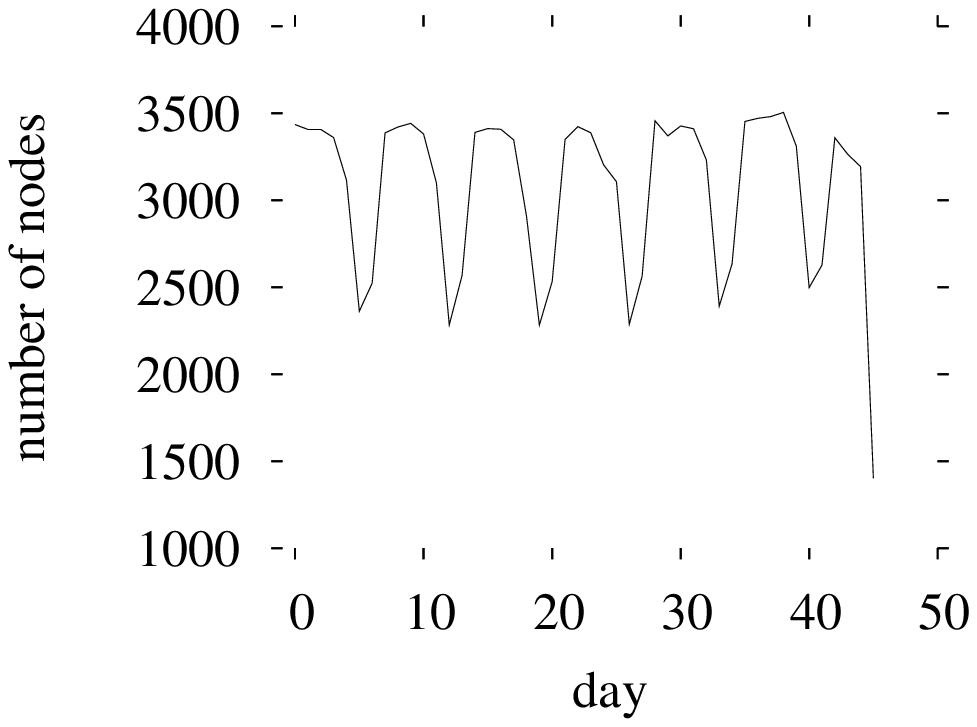}}
\subfigure[Locations visited]{\label{AP_pernode}\includegraphics[width=4cm]{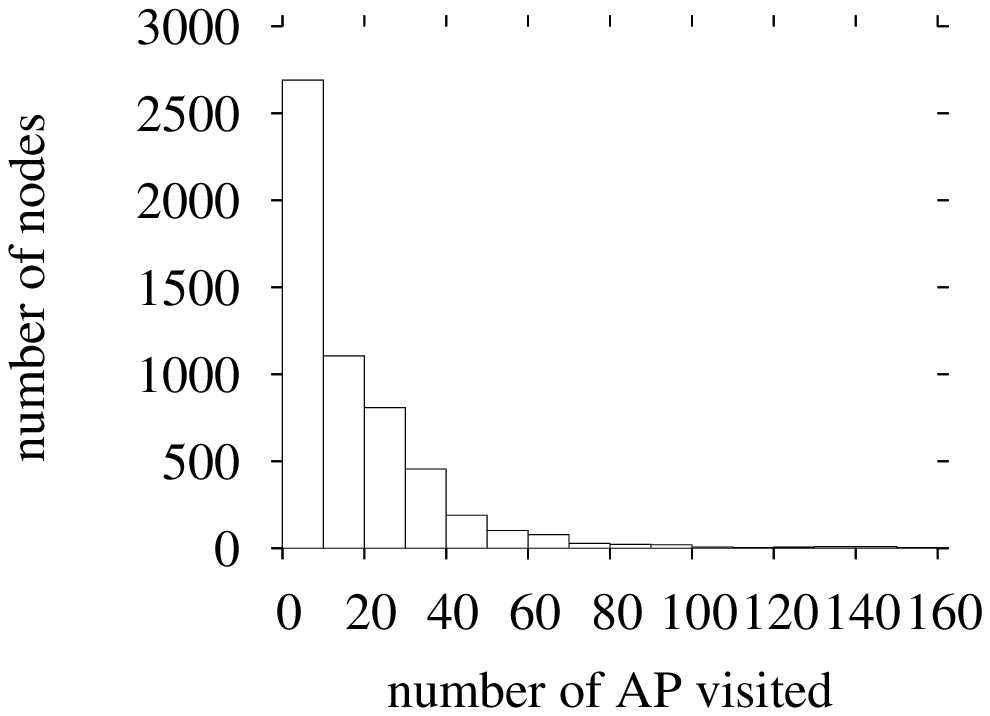}}
\subfigure[Locations visited per day]{\label{AP_perday}\includegraphics[width=4cm]{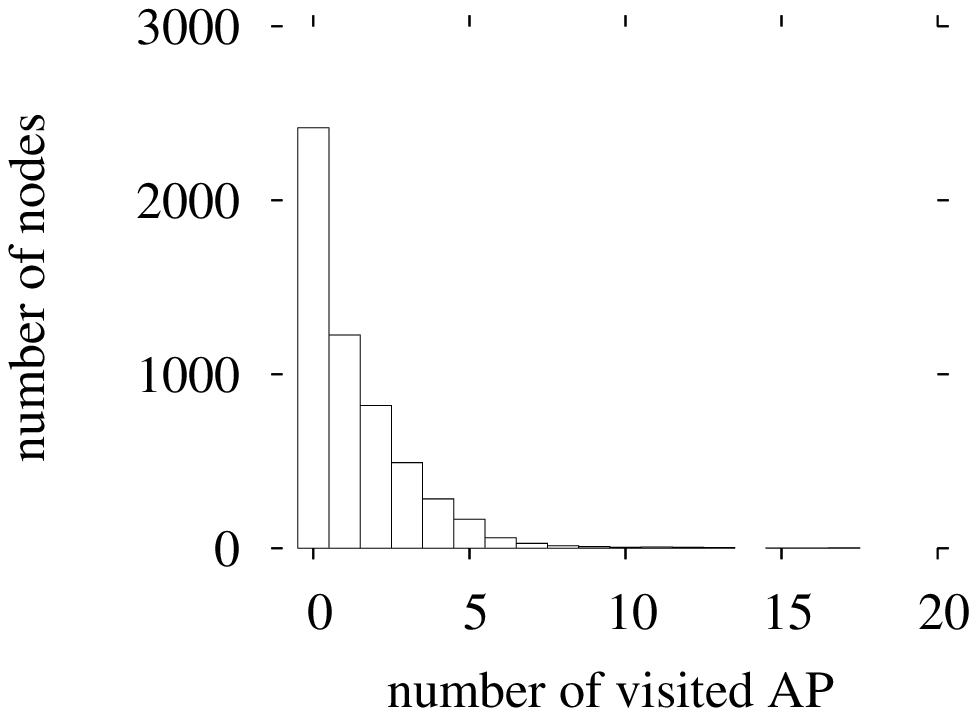}}
\subfigure[Connection time]{\label{time_pernode}\includegraphics[width=4cm]{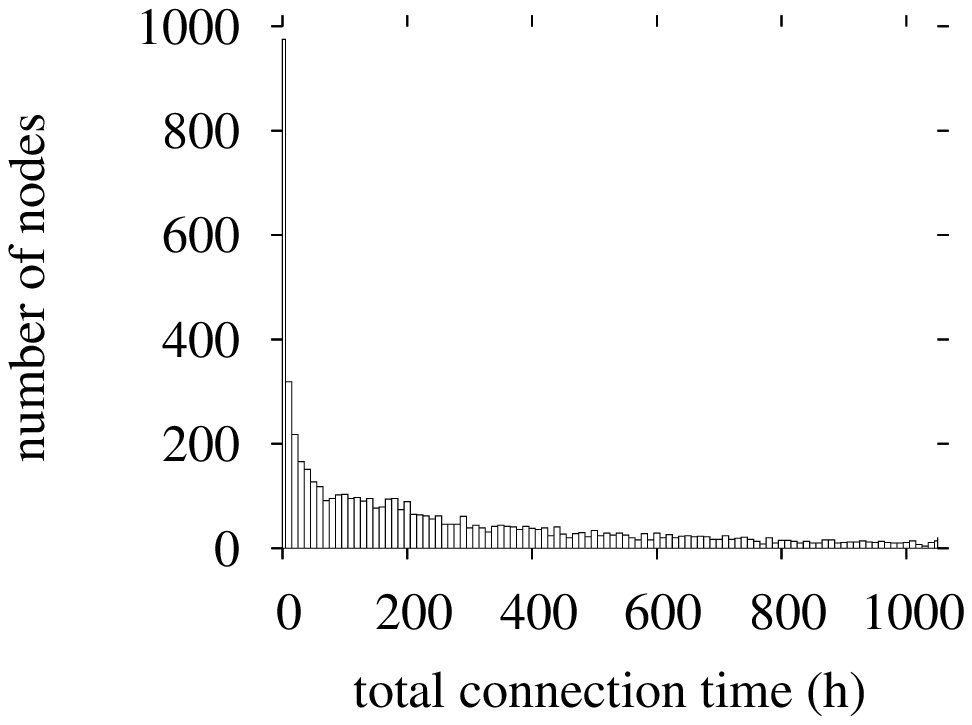}}
\subfigure[Connection time per day]{\label{time_perday}\includegraphics[width=4cm]{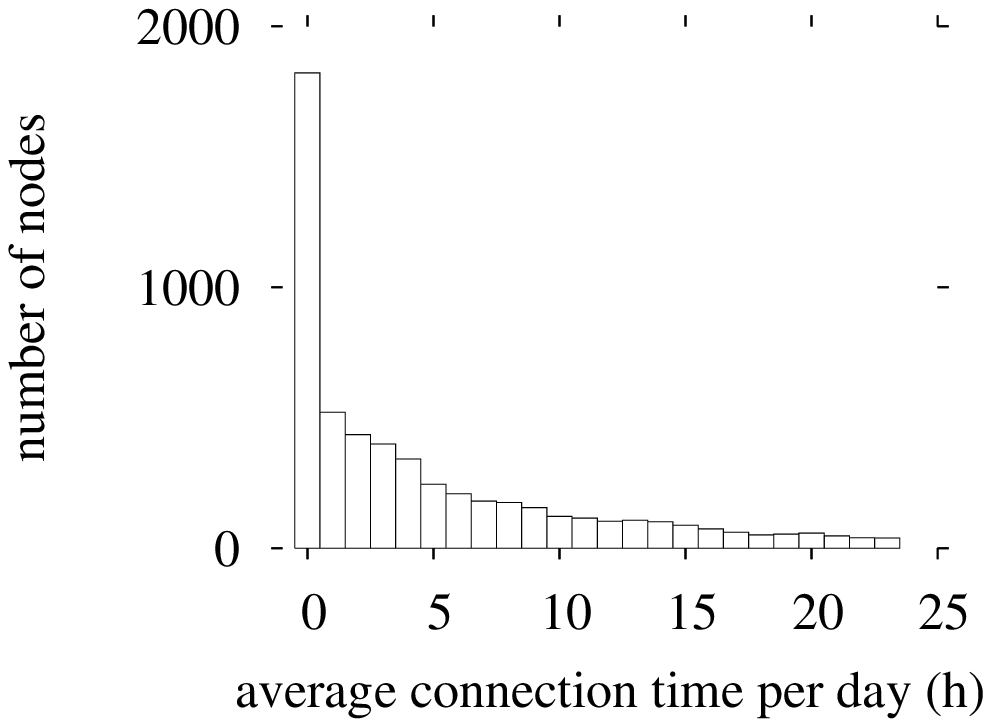}}
\subfigure[Apparition day]{\label{apparition}\includegraphics[width=4cm]{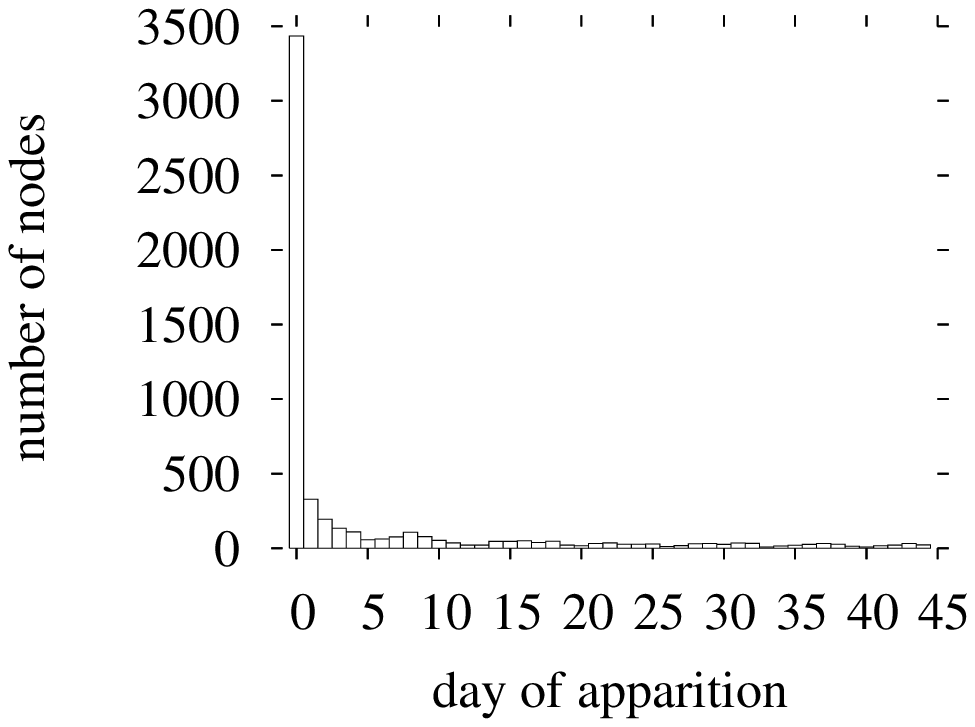}}
\subfigure[Disparition day]{\label{disparition}\includegraphics[width=4cm]{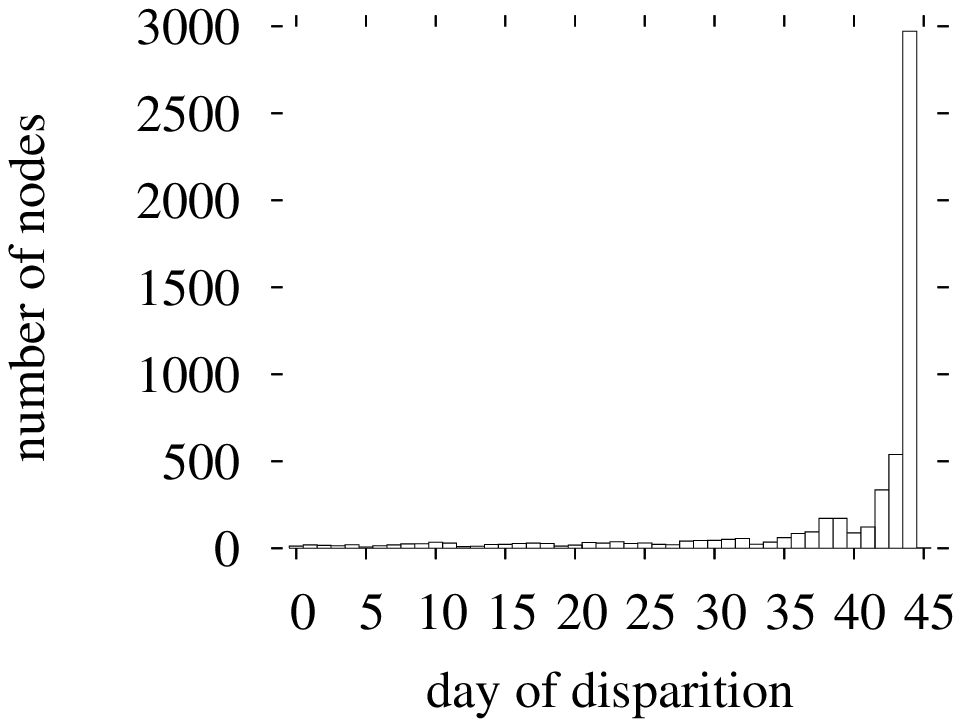}}
\caption{\label{fig_stats}
Statistics on the data set used for the simulations.}
\end{figure}

We replayed the mobility traces inferred from Dartmouth data between January \nth{26}
2004 and March \nth{11} 2004. Fig.~\ref{fig_stats} shows distributions 
that characterize users' behavior within this period. 
We choose that period because, as shown in
Fig.~\ref{activenodes_perday}, users make an intensive and regular use of the network.
As shown by Fig.~\ref{fig_activenodes}, this period is between Christmas 
and the spring break. In this period, we have observed a total of $5,545$ active users that have visited 
$536$ locations.

Users are mobile. They visit on average $16.66$ locations
in the period (see Fig.~\ref{AP_pernode}) and $1.75$ locations per day 
(see Fig.~\ref{AP_perday}). The distributions of the number of locations visited
by the nodes during the period and per day follow heavy tailed distributions. 
This means that 
the majority of users have a low level of mobility while some users are very 
mobile. Users with a low mobility level regarding the number of locations they visit may 
either be users that are not very present in the data or users that stay in one place, as in students who keep their 
laptop connected in their room at the student residence. 

The network usage displays a number of regularities. Fig.~\ref{activenodes_perday} 
shows the evolution of the number of active users 
per day. It highlights the existence of regular weekly cycles and a fairly constant 
number of active users: $2,901$ users per day on average. Regularity is 
a desirable property for this study
because we wish to evaluate the MobySpace based routing scheme in a context were
people move in their usual everyday environment having a number of constant habits.

Users make intensive use of the network. The mean presence time for that period is $243$ hours and is $5.18$ hours per day 
(see Fig.~\ref{time_perday}).  
Having users with a high level of presence is important but not sufficient. 
That presence must also be distributed over time. Thus, we analyse the 
distributions of the apparition and disparition days of users, and their total number of days
of presence. Fig.~\ref{apparition} and Fig.~\ref{disparition} show 
that apparitions and disparitions generally occur close to the limits of the 
period. This means that the probability that a node will disappear close to the beginning 
of the simulation is low. Similarly, the probability that a node will appear
for the first time close to the end of the period is low. 
Looking at the distribution of the number of days
that users are present (Fig.~\ref{activedays}), $25.48$ days on average, it appears that either users make an
intensive use of their laptop or PDA, or they seldom use it, but a majority 
of users make an intensive use of the network since $50\%$ of users are present more than 
$30$ days.

\subsection{Methodology}\label{sec_method}

We have implemented a stand alone simulator to evaluate the routing scheme. 
This simulator only implements the transport and network layers and it makes simple
assumptions regarding lower layers, for instance allowing for infinite
bandwidth. Nodes are supposed to have infinite buffers and are supposed to have the inherent 
knowledge of the mobility pattern of the others. We will study 
independently the problematic of learning mobility patterns in 
Sec.~\ref{sec_feasibility}.

We compare the performance of the MobySpace against the following:

\begin{itemize}
  
\item \textit{Epidemic routing}: This is 
  described by Vahdat and Becker~\cite{vahdat00epidemic}: Each time
  two nodes meet, they exchange their bundles. The major interest of 
  this algorithm is that it
  provides the optimum path and thus the minimum bundle delay.  We use
  it here as a lower bound.  
  This algorithm can be also seen as the extension of Dijkstra's shortest 
  path algorithm proposed  by Jain et al.~\cite{routingdtn} that takes into account 
  time-varying edge weights. In practice, epidemic routing suffers
  from high buffer occupancy and high bandwidth utilization.

\item \textit{Opportunistic routing}: A node waits to meet the destination in
  order to transfer its bundle. The main advantage of this method is
  that it involves only one transmission per bundle. 
  Bundle delivery relies just on the mobility of nodes and their
  contact opportunities.
  
\item \textit{Random routing}: There are many ways to define a random 
routing algorithm. In order to design one that acts similarly to the
MobySpace based routing scheme, we attribute for each destination node $j$ a 
preference list $l_j$, which is a randomly ordered list of all of the nodes.
When a node has a bundle destined to $j$, it sends
that bundle to the most preferred neighbor on the preference list $l_j$. If 
the most preferred neighbor has a lower preference than the current node,
the bundle is not forwarded.
This mechanism avoids loops by construction.

\item \textit{Hot potato routing}: When a node is at a location and the bundle's
  destination in not there, the node transfers the bundle to a
  neighbor chosen at random.  We have added a rule to avoid local
  loops: a node can only handle a bundle one time per location visit.

\end{itemize}

We will refer to them here by the following names: \emph{Epidemic}, using
Epidemic routing; \emph{Opportunistic}, using Opportunistic routing;
\emph{Random}, using Random routing; \emph{Potato}, using Hot potato 
routing, and \emph{MobySpace}, using the 
routing scheme that relies on the MobySpace.

All the scenarios share common parameters that can be found in Table~\ref{simulation_parameters}.  
We considered the whole set of $536$ locations that were visited over the course of the $45$ days of data. 
The virtual space used for routing
thus has $536$ dimensions. Due to the difficulty of running simulations with the totality 
of the $5,545$ nodes, especially with Epidemic, for which computation explodes with the number of nodes 
and the number of bundles generated, we used a sampling method. 
We have defined two kind of users: \emph{active}, which generate traffic, and 
\emph{inactive}, which only participate in the routing effort.
Every active node establishes a connection towards $5$ other nodes. An active node sends one bundle 
per connection.
For active users, we chose only the ones that appear at least one time in the first 
week of the simulations in order to be able to study bundle propagation over an extended period.
In each run, we sampled $300$ users with $100$ of them generating traffic.
The simulator used a time step of $1$s.

\begin{table}[!h]
\begin{center}
\small
\begin{tabular}{|l|c|}
\hline
\textbf{Parameter} & \textbf{Value}\\
\hline
Total nodes & 5545 \\
Total locations & 536 \\
Users sampled & 300 \\
Users generating traffic & 100 \\
Simulation duration & 45 days \\
Connections per user & 5 \\
Bundles per connection & 1 \\
Time step & 1 s \\
\hline
\end{tabular}
\end{center}
\caption{\label{simulation_parameters}Simulation parameters.}
\end{table}

We performed $5$ runs for each scenario. Simulation results
reported in the following tables are mean results with confidence intervals
at the $90$\% confidence level, obtained using the Student $t$
distribution.

\subsection{Results}\label{sec_res}

We evaluate the routing algorithms with respect to their transport layer performance. 
We consider a good algorithm to be one that yields a
low average bundle delay, the highest bundle delivery ratio and a low 
average route length. 

We consider two different kind of scenarios. One with only randomly sampled users and
one with only the most active.

\subsubsection{With randomly sampled users}$  $

In this scenario, we picked $300$ users completely at random and we replayed
their traces while simulating DTN routing. 

Table~\ref{res_avgresults} shows the average simulation results.
It shows for each of the implemented algorithms the average
bundle delay in number of days, the average delivery ratio, which corresponds to
the number of bundles received over the number of bundles sent, and the average route
length in number of hops. 

\begin{table}[!h]
\begin{center}
\small
\begin{tabular}{|c|c|c|c|c|c|c|}
\hline
  & \textbf{delivery ratio} & \textbf{delay} & \textbf{route length} \\
  & (\%) & (days) & (hops)\\
\hline
\textbf{Epidemic} & \co{82.0}{2.7} & \co{12.5}{0.9} & \co{7.10}{0.2} \\
\textbf{Opportunistic} & \co{4.9}{0.6} & \co{15.9}{2.5} & \co{1.0}{0.0} \\
\textbf{Random} & \co{7.2}{0.5} & \co{16.6}{2.6} & \co{3.12}{0.2} \\
\textbf{Potato} & \co{10.7}{1.7} & \co{19.1}{1.6} & \co{72.7}{16.5} \\
\textbf{MobySpace} & \co{14.9}{2.9} & \co{18.9}{1.0} & \co{3.8}{0.2} \\
\hline
\end{tabular}
\end{center}
\caption{\label{res_avgresults}Results with randomly sampled users.}
\end{table}

The first thing we can observe is the fact that within the 45 days of simulation there is
still a certain number of bundles that are not delivered with Epidemic. The mobility of the
$300$ nodes or their level of presence were not sufficient to ensure all the deliveries. 
Our sample included just $5\%$ of the entire set of nodes. 
By deploying this system on more nodes, the delivery ratio would rise closer to $100\%$.
Furthermore, we did not select nodes based on their mobility characteristics.
Some of the nodes may have poor mobility.

\begin{figure}[!h]
\centering
\subfigure[Epidemic]{\includegraphics[width=2.7cm]{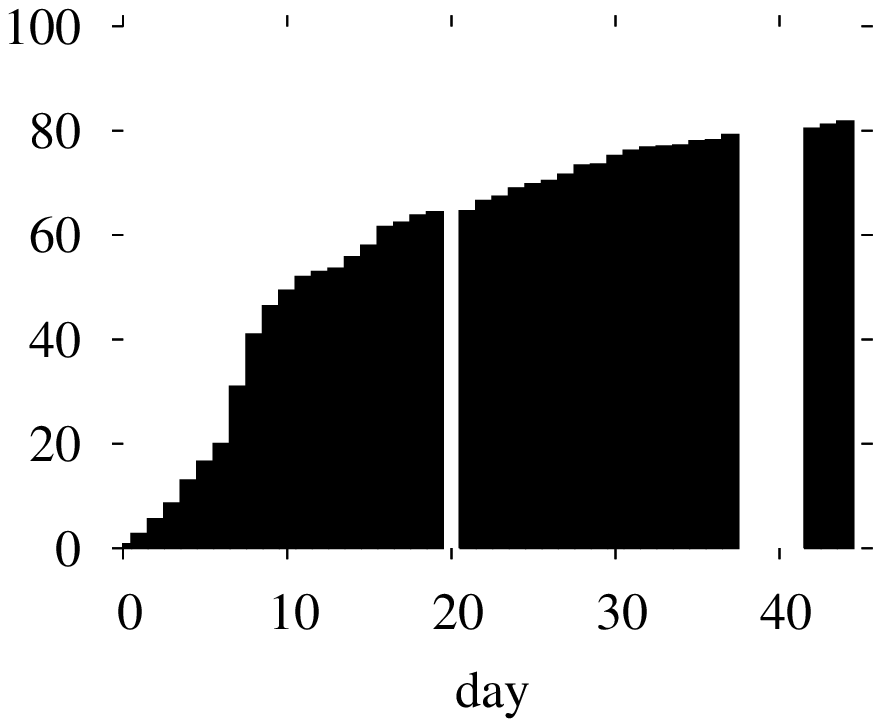}}
\subfigure[Opportunistic]{\includegraphics[width=2.7cm]{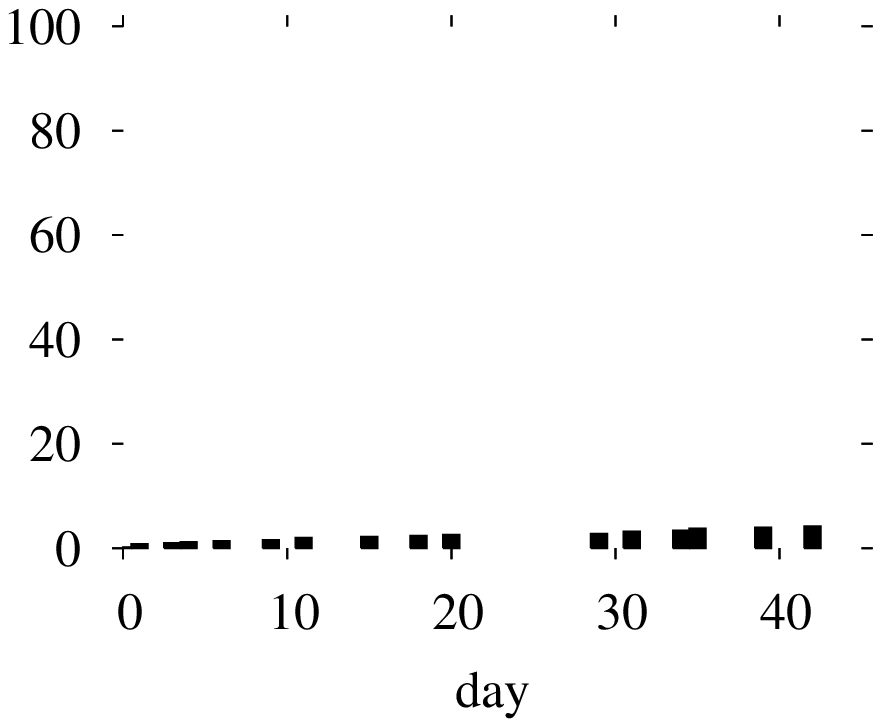}}
\subfigure[Random]{\includegraphics[width=2.7cm]{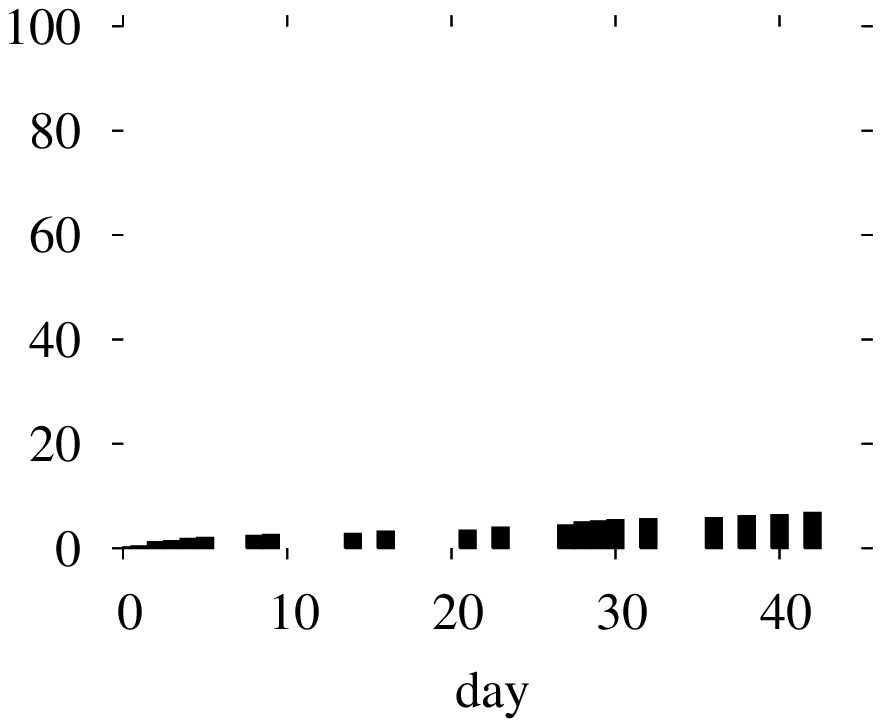}}
\subfigure[Potato]{\includegraphics[width=2.7cm]{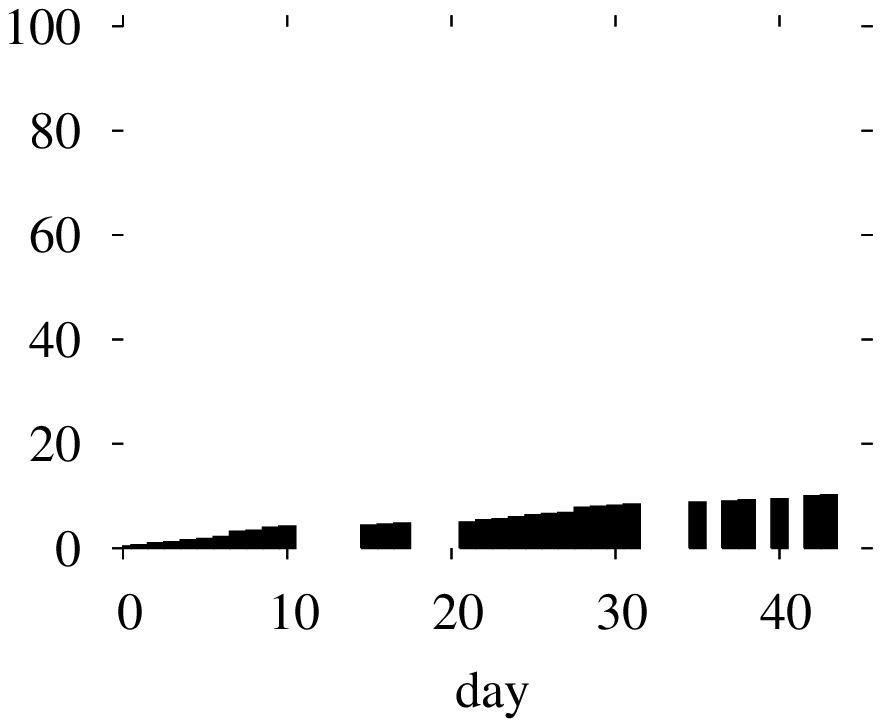}}
\subfigure[MobySpace]{\includegraphics[width=2.7cm]{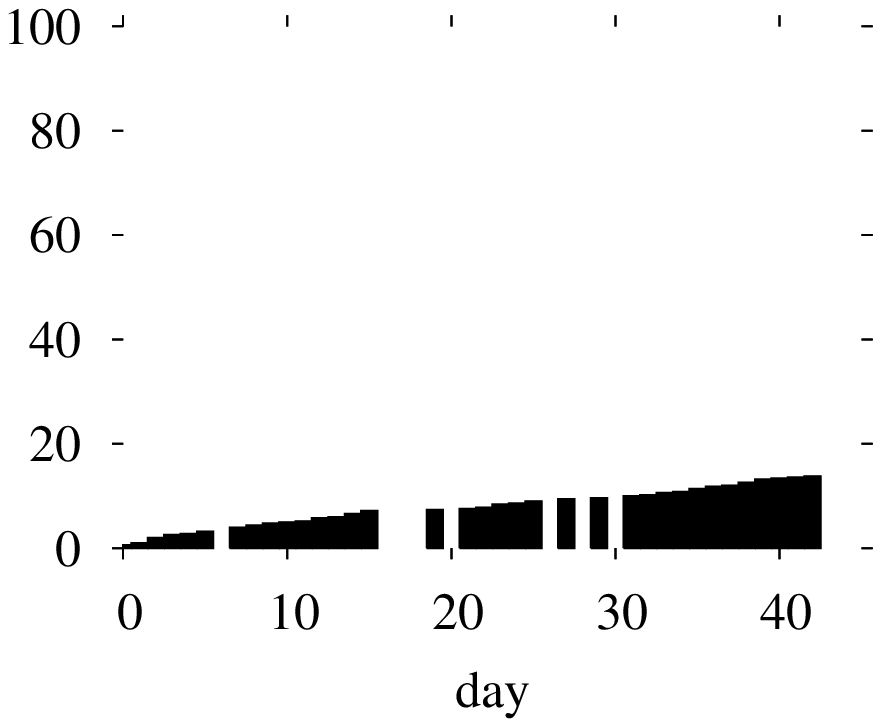}}
\caption{\label{res_avgresults.cdf}Cumulative distribution of packets delivered over time.}
\end{figure}

Table~\ref{res_avgresults} shows that MobySpace delivers twice as many bundles as Random 
but still far less than Epidemic, which does not miss any opportunities. Random delivers 
somewhat more bundles than Opportunistic because the bundles are more mobile. This phenomenon is 
even true for Potato, which outperforms Random but delivers fewer bundles than MobySpace. 
At first glance, the average bundle delay of MobySpace seems poor. We believe this average is
influenced by the fact that more bundles are delivered compared to the other schemes, except Epidemic. 
The additional bundles delivered by MobySpace might be more difficult to route than 
the others, leading to higher delays. The investigation of this issues is kept for
future work.
However, the average bundle delay is an interesting indicator of the performance
an algorithm can achieve.
Fig.~\ref{res_avgresults.cdf} presents the cumulative distribution of 
packets delivered over time. It shows why the average bundle delay is higher for
MobySpace compared to Random. It is simply because MobySpace delivers more packets in 
a constant fashion over time.
Looking now at the average route lengths,
we see that in all the cases, except Potato, they are lower than for Epidemic. MobySpace
engenders routes that are about half as long as those created by Epidemic. With MobySpace, bundles are 
transmitted from a node to another because of their mobility patterns, not simply 
because of the opportunities of contact. 
Potato engenders routes that are extremely long because, at each contact, bundles switch from 
one node to another. Potato may not be suitable for a real system because of 
bandwidth and energy consumption issues. 

\subsubsection{With the most active users}$  $

We also evaluate routing in a scenario with only the most active users, to see the effect of activity on 
performance. Such a scenario might also be more typical of an ambient network environment.
Several metrics can characterize the level of activity.
We use the regularity of the users' presence in the network, 
as measured by the number of active days. 
The number of users in our data that are active all $45$ days is $835$. 
We consider these users as a pool from which we sample for each 
simulation run.

\begin{table}[!h]
\begin{center}
\small
\begin{tabular}{|c|c|c|c|}
\hline
  & \textbf{delivery ratio} & \textbf{delay} & \textbf{route length} \\
  & (\%) & (days) & (hops)\\
\hline
\textbf{Epidemic} & \co{96.7}{1.9} & \co{3.1}{0.4} & \co{7.9}{0.3} \\
\textbf{Opportunistic} & \co{10.7}{1.1}& \co{17.6}{1.6} & \co{1.0}{0.0} \\
\textbf{Random} & \co{14.0}{1.0}& \co{17.9}{1.8} & \co{3.5}{0.1} \\
\textbf{Potato} & \co{38.9}{1.0}& \co{19.1}{0.4} & \co{317.0}{29.0} \\
\textbf{MobySpace} & \co{50.4}{4.7} & \co{19.5}{1.3} & \co{5.1}{0.2} \\
\hline
\end{tabular}
\end{center}
\caption{\label{res_avgresults_mostactive}Results with the most active users.}
\end{table}

Table~\ref{res_avgresults_mostactive} shows the average simulation results.
Considering only the most active users, more bundles are delivered by the algorithms. 
MobySpace attains a delivery ratio of 50.4\% instead of 14.9\%.
The delivery ratio of MobySpace would have been, as previously, higher if more nodes 
had participated in the scenario. We intend to study this in future work
by performing larger simulations.
The average bundle delay achieved is very low for Epidemic compared to the other algorithms.
Since nodes are more present in the network, Epidemic certainly needs fewer relays to 
deliver the packets. Route lengths are still less than Epidemic for Opportunistic,
Random, and MobySpace, whereas it is higher for Potato compared to the previous
scenario with randomly sample users. 

These results confirm that the MobySpace evaluated in this paper enhances routing as compared 
to various generic approaches for 
routing in an ambient network formed by users carrying personal devices in a campus setting. 
MobySpace achieves a high delivery ratio compared the simple algorithms like Opportunistic, Random, or Potato. 
It also leads to low bandwidth usage by using short route compared to Epidemic.  

\section{Feasibility}\label{sec_feasibility}

Sec.~\ref{sec_res} has shown encouraging results for the use of MobySpace. However, the 
simulations rely on the assumption that nodes are aware of their mobility 
patterns. This section examines two different factors 
that impact the feasibility of this architecture: the characteristics of the 
mobility patterns and the possibility of learning them.

\subsection{Mobility pattern characteristics}

As noted in our prior work~\cite{wdtn}, when nodes do not have a high degree
of segregation in their mobility patterns, MobySpace can not  
benefit from the patterns for efficient routing. We analyse here the properties of 
the mobility patterns we compute on users of Dartmouth College with the help of 
the relative entropy, $S_r$, applied to the set of probabilities that make up a mobility pattern. 
This metric describes the homogeneity of mobility patterns, which is $1$ for a
pattern with no preference among locations and is small for patterns that strongly 
prefer a few locations. It is defined by:

\begin{equation}
S_r=- \frac{ \sum_ {i=1}^{n} p_i \ln p_i } {\ln n}
\end{equation}

The relative entropy is relevant for the analysis of mobility patterns
because it captures a number of important characteristics. 
The relative entropy is at the same time correlated to
the number of locations visited and to the time spent at each location.
If a node is equally likely to be found in any location, it has the
maximum relative entropy value of 1.  If it is very likely to be
found in one of a few locations, and unlikely to be found in any
other, it has low relative entropy.

\begin{figure}[!h]
\begin{center}
\includegraphics[width=6cm]{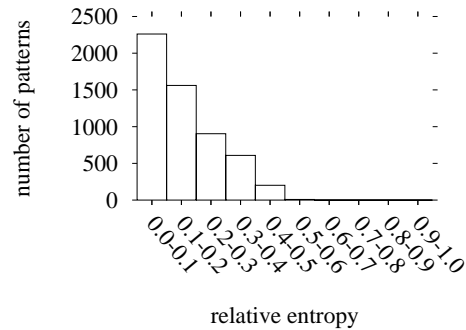}
\caption{\label{fig_entropie} Relative entropy distribution of mobility patterns.}
\end{center}
\end{figure}

Fig.~\ref{fig_entropie} shows the distribution of the relative entropy of users' mobility 
patterns for the period of $45$ days. They display generally  
low entropy: on average $0.15$. The patterns tend to demonstrate good properties
for the MobySpace routing scheme because either they contain few components 
or they contain many components in a non homogeneous fashion.  

We study the effect of pattern entropy on MobySpace routing.
Table~\ref{res_entropy} shows that the relative entropy of mobility patterns has a great influence
the performance in terms of the number of packets that are delivered.
The higher the relative entropy, the higher the delivery ratio.
Route lengths are stable over the increase of the relative entropy, except for Potato
that generates longer routes.  

These results show that a lack of diversity in the movements of users does not 
favor routing in such an environment. In our prior work~\cite{wdtn}
we demonstrate, with an artificial scenario, that too much diversity can also be a problem
if mobility patterns can not be distinguished. In that case, distances in MobySpace have little significance. 
We were not able to 
reproduce this demonstration with Dartmouth data because there is no user in the data 
that visits almost all the locations in a regular fashion. 
We can conclude that a MobySpace approach is of interest when mobility patterns display 
a low relative entropy, but not too close to $0$. 

\begin{table}[!h]
\begin{center}
\scriptsize
\begin{tabular}{|c|c|c|c|c|}
\hline
metric & $S_r$ & \textbf{delivery ratio} & \textbf{delay} & \textbf{route lengths} \\
  & & (\%) & (days) & (hops)\\
\hline
\textbf{Epidemic} & $[0.0-0.1]$ & \co{45.4}{5.1} & \co{24.1}{1.7} &  \co{7.0}{0.2} \\
& $[0.1-0.2]$ & \co{79.6}{3.2} & \co{13.1}{1.8} & \co{8.0}{0.4} \\
& $[0.2-0.3]$ & \co{97.8}{1.7} & \co{8.7}{1.3} & \co{7.5}{0.4} \\
& $[0.3-0.4]$ & \co{99.0}{0.5} & \co{6.0}{0.9} & \co{7.1}{0.4} \\
\hline
\textbf{Opportunistic} & $[0.0-0.1]$ & \co{2.2}{0.3} & \co{15.0}{3.8} & \co{1.0}{0.0} \\
& $[0.1-0.2]$ & \co{4.4}{0.9} & \co{19.8}{2.4} & \co{1.0}{0.0} \\
& $[0.2-0.3]$ & \co{9.6}{2.0} & \co{19.9}{1.0} & \co{1.0}{0.0} \\
& $[0.3-0.4]$ & \co{24.5}{2.5} & \co{10.9}{0.9} & \co{1.0}{0.0} \\
\hline
\textbf{Random} & $[0.0-0.1]$ & \co{2.3}{0.4}  & \co{11.6}{4.5}  & \co{2.0}{0.3} \\
& $[0.1-0.2]$ & \co{5.8}{1.2} & \co{20.0}{2.6} & \co{3.0}{0.2} \\
& $[0.2-0.3]$ & \co{12.3}{1.4} & \co{17.6}{2.5} & \co{3.5}{0.1} \\
& $[0.3-0.4]$ & \co{29.5}{3.0} & \co{12.5}{1.1} & \co{3.9}{0.1} \\
\hline
\textbf{Potato} & $[0.0-0.1]$ & \co{3.2}{0.8}  & \co{16.9}{1.4}  & \co{43.0}{12.0} \\
& $[0.1-0.2]$ & \co{9.6}{1.1} & \co{19.8}{2.8} & \co{116.2}{44.2} \\
& $[0.2-0.3]$ & \co{19.8}{5.6} & \co{20.2}{1.5} & \co{162.7}{44.7} \\
& $[0.3-0.4]$ & \co{36.6}{4.9} & \co{12.0}{1.3} & \co{176.6}{14.3} \\
\hline
\textbf{MobySpace} & $[0.0-0.1]$ & \co{3.4}{0.4}  & \co{14.9}{1.8}  & \co{2.5}{0.2} \\
& $[0.1-0.2]$ & \co{8.4}{2.4} & \co{19.5}{2.3} & \co{3.3}{0.2} \\
& $[0.2-0.3]$ & \co{19.8}{2.4} & \co{19.7}{1.2} & \co{4.0}{0.2} \\
& $[0.3-0.4]$ & \co{42.3}{4.8} & \co{13.4}{1.3} & \co{4.7}{0.2} \\
\hline
\end{tabular}
\end{center}
\caption{\label{res_entropy}Results with users having different entropy.}
\end{table}

\subsection{Space reduction}

Because transmitting nodes' mobility patterns can potentially consume bandwidth,
we evaluate a scenario in which nodes only diffuse the main components of their mobility patterns. 
We ran simulations taking into account only the
principal \nth{1}, \nth{2}, or \nth{3} components of
mobility patterns of nodes, and we consider the most active users.

\begin{table}[!h]
\begin{center}
\small
\begin{tabular}{|c|c|c|c|c|}
\hline
 l & \textbf{delivery ratio} & \textbf{delay} & \textbf{route length} \\
  & (\%) & (days) & (hops)\\
\hline
 $l=1$ & \co{39.2}{5.9} & \co{20.2}{2.6} &  \co{4.9}{0.4} \\
 $l=2$ & \co{46.3}{3.3} & \co{19.9}{1.2} & \co{5.2}{0.2} \\
 $l=3$ & \co{47.5}{4.6} & \co{19.4}{1.8} & \co{5.2}{0.2} \\
 $l=536$ & \co{50.4}{4.7} & \co{19.5}{1.3} & \co{5.1}{0.2} \\ 
\hline
\end{tabular}
\end{center}
\caption{\label{res_partial}Results with space reduction. $l$ is the number of components taken into account.}
\end{table}

Table~\ref{res_partial} shows that the higher the number of components taken into account, 
the higher the performance. Surprisingly, the delivery ratio tends very quickly to that of 
the scenario where all the components are used. These simulations show that 
only few components are needed to be exchanged between nodes in order to 
perform routing. 

\subsection{Mobility pattern learning}

One important condition for the applicability of the MobySpace 
is whether users can learn their own mobility patterns. In this section we provide a first 
study on this issue with the Dartmouth data.

For that purpose, we split the $45$ days of Dartmouth data into two periods:
the learning period and the routing period. 
The learning period consists of the first $15$ days and the routing period, the last 
$30$ days.
We study here how well the mobility patterns 
of nodes learnt in the learning period match the mobility patterns 
that characterize the routing period. The error is measured as to be the Euclidean distance 
between the two mobility patterns, divided 
by the maximum possible distance between two mobility patterns in the hyperplane:

\begin{equation}
e=\frac{d_{ij}}{\sqrt n}\textrm{, with $n$ the number of dimensions}
\end{equation}

\begin{figure}[!h]
\begin{center}
\includegraphics[width=7cm]{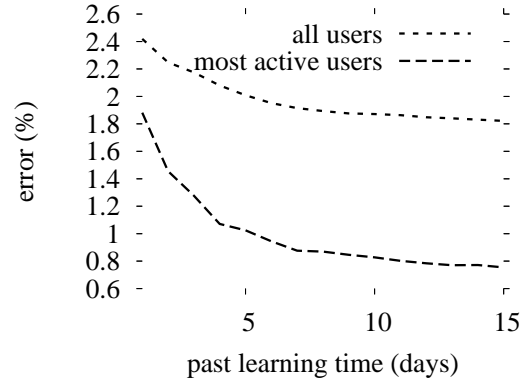}
\caption{\label{fig_avgerrors_vs_day} Prediction error of mobility patterns.}
\end{center}
\end{figure}

We varied the number of days devoted to learning during the learning
period, starting with the one day immediately prior to the routing
period, and working back to cover all 15 days of the learning period.
Fig.~\ref{fig_avgerrors_vs_day} shows the prediction error of mobility patterns, as a function 
of the number of days devoted to learning. We made this computation
for all the nodes and for only the most active ones. We see that, in both cases,
the longer nodes learn their own mobility, the closer their mobility patterns 
approximate the patterns of the routing period. As expected, the most active users learn their patterns more rapidly 
than the others.

These initial results on the ability of nodes to learn their own mobility patterns 
are encouraging. They indicate that nodes might be able to benefit from their past knowledge to make 
routing decisions within the MobySpace. Nevertheless, further studies are needed to quantify 
possible long and short term dependencies in mobility traces. This must be also validated on other 
mobility traces. 

\section{Related work}\label{related}

Some work concerning routing in DTNs has been performed
with scheduled contacts, such as the paper by
Jain et al.~\cite{routingdtn} that tries to improve the connectivity
of an isolated village to the internet based on knowledge of when a
low-earth orbiting relay satellite and a motor bike might be available
to make the necessary connections. Also of interest, work on
interplanetary networking~\cite{ipn,ipnbis} uses predicted contacts such as
the ones between planets within the framework of a DTN architecture.

The case of only opportunistic contacts has been analyzed by Vahdat
and Becker~\cite{vahdat00epidemic} using the epidemic routing scheme that consists of flooding.
The ZebraNet project~\cite{zebranet} is exploring this idea to perform studies of animal
migrations and inter-species interactions. Data are flooded in the network such as
they get back to access points using animals' mobility.
In order to control flooding in DTN, Spyropoulos et al.\ have introduced
the Spray and Wait \cite{spray} protocol that distributes a number of copies
to relays and then waits until the destination meets one of them.
Harras et al.~\cite{harras_networking05} have evaluated simple controlled message flooding schemes
with heuristics based, for instance, on hop limits or timeouts. They also introduce a
mechanism based on packet erasure.  Once a message arrives at the destination
after basic flooding, the remaining copies in the buffers of other nodes are erased.
Wang et al.~\cite{wang_wdtn} reincode the messages with erasure codes and distribute
their different parts over a large number of relays, so that the original messages
can be reconstituted even if not all packets are received. Widmer et
al.~\cite{widmer_wdtn} have explored network coding techniques.
All these approaches distribute multiple copies of packets, they ensure a high reliability
of delivery, and a low latency, but they imply high buffer occupancy and high bandwidth consumption.
Small et al.~\cite{small_wdtn} propose an analytical study of existing trade-offs between
resources consumption such as energy, throughput, buffers and the
performance in term of latency.

Some research projects such as Data Mules~\cite{datamules} or
SeNTD~\cite{sendt} use mobile network elements to transport data from fixed sensors
to a number of access points in an opportunistic fashion. For instance, in SeNTD,
data from sensors placed on buoys that monitor
the water quality on a lake are relayed by tourist tour-boats or pleasure cruisers.

A large amount of work concerning routing in DTNs has also been performed with
predicted contacts, such as the algorithm of Lindgren et
al.~\cite{lindgren03}, which relies on nodes having a community
mobility pattern. Nodes mainly remain inside their community and
sometimes visit the others.  As a consequence, a node may transfer a
bundle to a node that belongs to the same community as the
destination. This algorithm has been designed as a possible solution to provide
internet connectivity to the Saami~\cite{saami} population who live in Swedish Lapland
with a yearly cycle dictated by the natural behavior of reindeer.
In a similar manner, Burns et al.~\cite{mvrouting} propose
a routing algorithm that uses past frequencies of contacts.
Also making use of past contacts, Davis et al.~\cite{davis01} improved
the basic epidemic scheme with the introduction of adaptive dropping
policies. Recently, Musolesi et al.~\cite{musolesi05} have introduced
a generic method that uses Kalman filters to combine and evaluate the
multiple dimensions of the context in which nodes are in order to take routing decisions.
The context is made of measurements that nodes perform periodically,
which can be related to connectivity, but not necessarily.
This mechanism allows network architects to define their own hierarchy
among the different context attributes. LeBrun et al.~\cite{lebrun} propose
a routing algorithm for vehicular DTNs using current position and
trajectories of nodes to predict their future distance to the destination. 
They replay GPS data collected
from actual buses in the San Francisco MUNI System, through the NextBus
project. Finally, Jones et al.~\cite{jones_wdtn} propose a
link state routing protocol for DTNs that uses the minimum expected delay
as the metric.

\section{Conclusions and future work}\label{sec_conclu}

The main contribution of this paper has been the validation of a
generic routing scheme that uses the formalism of a high-dimensional
Euclidean space constructed upon mobility patterns, the MobySpace.  We have shown
through the replay of real mobility traces that it can applied to DTNs
and that it can bring benefits in terms of enhanced bundle delivery and
reduced communication costs. 

This paper has also presented results of a feasibility study in order to 
determine the impact of the characteristics of nodes' mobility patterns on 
the performance and to study nodes' ability to learn their patterns.
Thus, to make DTN routing work with the MobySpace, nodes need to have a minimum level 
of mobility with mobility patterns that can be sufficiently discriminated. 
We present encouraging results about
the capacity of nodes to learn their own patterns.
And, we also see that nodes can reduce 
the number of components in the mobility patterns without 
a high impact on routing performance. This can reduce the overhead of MobySpace and 
the complexity of handling mobility patterns.

Future work along these lines might include studies concerning the
impact of the structure of the Euclidean space, i.e., the number and
type of dimensions, and the similarity function.  Different kind of
Euclidean space have to be investigated by considering schemes like
the one described in Sec.~\ref{sec_freq_of_visit_MobySpace} that takes for each dimension
the frequency of contacts between a certain pair of nodes or the one 
that captures frequential properties during nodes' visits to locations.  

Further work remains to be done on the stability of mobility patterns over time
and their ability to be learned by nodes. The patterns may contain long term and short 
term dependencies. Nodes can have different mobility patterns that are each stable. 
For instance, they can have one for the week-ends, one for the vacations, and one for 
working weeks. 

Additionnaly, further validations need to be conducted on real data and in different environments.
MobySpace can be tested on traces coming from larger cell networks, like
GSM networks. We might also want to evaluate MobySpace in different social contexts where
nodes have specific mobility patterns.

\section*{Acknowledgments}

We gratefully acknowledge David Kotz for enabling our use of wireless
trace data from the CRAWDAD archive at Dartmouth College. We
thank Marc Giusti and Pierre Lafon at the STIX laboratory (\'Ecole
Polytechnique / CNRS) for access to the machines we used for the
simulations. This work was supported by E-NEXT,
an FP6 IST Network of Excellence funded by the European Commission.
Also, LiP6 and Thales Communications, which supported this work through their joint research
laboratory, Euronetlab, and the ANRT (Association Nationale de la
Recherche Technique), which provided the CIFRE grant 135/2004.

\bibliographystyle{IEEE}

\end{document}